# Boolean Algebraic Programs as a Methodology for Symbolically Demonstrating Lower and Upper Bounds of Algorithms and Determinism

Daniel J. McCormack

June 24,2014

## Abstract

The lower and upper bound of any given algorithm is one of the most crucial pieces of information needed when evaluating the computational effectiveness for said algorithm. Here a novel method of Boolean Algebraic Programming for symbolic manipulation of Machines, Functions, and Inputs is presented which allows for direct analysis of time complexities and proof of deterministic methodologies. It is demonstrated through the analysis of a particular problem which is proven and solved through the application of  Boolean algebraic programming.

## 1|Problem Setup

Any problem of a language A is an element of P(Polynomial class) iff it can be transformed into and output O by a machine M within a time bounded by the Polynomial S(x) where x is the length of A.[1] If O can be checked by a machine M' within a polynomial S'(x) to satisfy the rules of A via a non-deterministic way such that given $(1.0) \forall O \exists R: (R \in \rho) \neq (R_t \in \rho)$ where R is the steps taken by machine and $\rho$ is the set of steps for all runs of the machine.[2]

## 2|Boolean Satisfiability Problem(BSAT)

The Boolean Satisfiability problem asks whether there is a way to evaluate if a Boolean algebraic expression is satisfiable (TRUE) without testing each possible combination of the variables. The Boolean Satisfiability was the first problem proven to be NP- Complete by the Cook-Levin theorem. The theorem states $C \in NPC$ iff C=BSAT.

## 3|Contextualization

The P Versus NP problem is a question related directly to mathematics and computer science. This problem essentially asks whether methodologies exist for solving traditionally computationally intensive problems through deterministic algorithms that have polynomial run times. Suppose we are given the following situation, which is a generalization of the problem stated by the CMI:

Sort a list S into a Nx2 matrix O such that in any row r of O no two members of any given r are also members of the same row of I, a list containing an even integer which is less than or equal to the number of elements in S.[3]

## 4|Non Deterministic Solution in Polynomial Time

**Notation**

When describing Functional Operations I will frequently use <Machine | Operator[*]Language> as the format for describing Turing Machine operations. Language Simply means the input. *implies that the preceding text is an operator.

**Theorem**

The Problem can be checked non deterministically in polynomial time or is of class NP

**Definition**

Our Language Y is of finite length and K is simply a Boolean expression or set of expressions which are satisfiable for some particular State O.

**Proof**

Let us call our input language Y. Suppose our Algorithm is called V and that our criteria is K. V:

$$O \exists : \, < M | V^* Y > \rightarrow O \in K$$

Meaning that Machine M will perform an operation $V^*$ upon Y which produces O, an output, that satisfies K.

## 5|Reduction to the Boolean Satisfiability (B-SAT) Problem

Any NP problem is NP complete if it can be reduced to the (B-SAT) Problem. For the selected problem it can be shown through a direct proof:

**The SAT Reducibility Theorem**

Given an array S of an arbitrary number of elements, where each element contains a property F, which denotes elements with which it cannot be paired: $F_x = Y \therefore \{X, Y\} \notin D_n$, and any element X cannot be used twice, the task of sorting can be stated as a variation of the Boolean Satisfiability Problem.

**Definition A**

For an element X within each row of D, the second element of D cannot be an element of the list of incompatibles $I_X$. Furthermore, for a given Element Y, which is the second element of D, $X \notin I_Y$. For practical purpose, the Incompatibles must only be checked once because:

If $X \notin I_Y$ then $Y \notin I_X$ and the corollary is **TRUE** by definition of the problem.

**Definition B**

D is composed of an arbitrary number of elements of S. For our purposes sets of two X and Y will be used within the Boolean Algebraic Equations. $I_n$ is the current set of Incompatibles which is defined by the properties of the first element of each row of D.

**Definition C**

Each row of D room contains two Elements X,Y and thus:

$$(\{X,Y\} \in D_a) \,\overline{\wedge}\, (\{X,Y\} \in D_b) \,\overline{\wedge}\, \dots (\{X,Y\} \in D_n)$$

**Definition D**

The Sorted list of Elements within all instances of D is referred to as K.

**Proof:**

In order to find the sorted list K which contains a solution in which no incompatibles are listed together we write the following Boolean expression:

$$\forall K \exists K \colon \left(X_1 \notin I_{Y_1} \wedge Y_1 \notin I_{X_1}\right) \wedge \left(X_2 \notin I_{Y_2} \wedge I \notin I_{X_2}\right) \dots$$

In order to check that elements are not utilized more than once:

$$\forall X \nexists K : (X \in D_a) \wedge (X \in D_{a \pm n}), n \neq 0$$

# 6|An Introduction to Algorithmic Time Complexity

Time complexity of Algorithms is the analysis of the maximal length of time an algorithm could require to compute an answer given an input input of length n. The system for denoting linear time complexity would be O(n), meaning that the upper time boundary for the function is based on the size (n) of the input . A more general case is polynomial time, abbreviated as P. A P class Algorithm means that given any size input, the maxiumu length of time required to provide an answer is something of the form:

$$(1) O(n^k)$$

where k is an arbitrary constant. The standard algorithm for the addition of i length numbers is O(n) because there is one step for each integer. Multiplication however, is of the form O(n$^2$) because there is the act of multiplication, and then the act of addition if i>1 for a given multiplier. The form of Algorithm, either deterministic or non-deterministic is also of exceptional interest. Throughout this paper deterministic polynomial time algorithms will be represented as P. Nondeterministic polynomial Algorithms will simply be denoted as NP.

NP algorithms, do not calculate answers in deterministic ways; instead steps taken by the Algorithm for a language L varies between runs. In other words, the algorithm does not always have identical "behavior" for the same input. The P Versus NP problem asks whether there are algorithms which are P class, which can solve NP problems. Solving this problem would encourage the search for algorithms which can deterministically provide answers to questions which could require the upper bound of the O(n$^k$) for a given input.

# 7|Motivation for Determinism

Suppose the list $D$ is generalized as follows:

$$(2) \begin{bmatrix} e_0 & e_1 \\ \cdots & \cdots \\ e_n & e_{n+1} \end{bmatrix}$$

then the total number of ways to arrange the S(e), assuming that there are no incompatible arrangements of S is: $e_{n+1}!$ . However, in the context of the dean problem there are many incompatibles. If there are two incompatibles the expression for the placement of them in a 2 by 1 matrix,$\begin{bmatrix} x & y \end{bmatrix}$ would be 2!. The number of incompatibles however could be greater than two, therefore the equation must factor in for more occurrences:

$$(3) occurences = n! \cdot i!$$

The Equation for occurrences assumes that the matrix is only large enough to hold n sets of incompatibles. Therefore occurrences is multiplied by $e_{n+1}$ however the number of rows available decreases as n increases therefore if n is greater than 2, Occurrences is multiplied by $e_{n+1} - \frac{a}{2}$, where $a_x = a_{x-1} + 2$ and x is the element in the series such as: $x_0 \ldots x_1 \ldots x_n$. However this will need to be a subtractive series, such that:

$$(4) \varphi = (e_{n+1})! - \left((2!\, i!)(e_{n+1})\right) - \left((4!\, i!)\left(e_{n+1} - \frac{4}{2}\right)\right) - \left((6!\, i!)\left(e_{n+1} - \frac{6}{2}\right)\right) \ldots$$

Or more generally:

$$(5) \varphi = (e_{n+1})! - \left((a!\, i!)(e_{n+1})\right) - \left(([a+2]!\, i!)\left(e_{n+1} - \frac{[a+2]}{2}\right)\right) - \left(([a+4]!\, i!)\left(e_{n+1} - \frac{[a+4]}{2}\right)\right) \ldots$$

if a =2. The series ends when $e_{n+1} - \frac{a}{2} = 0$. The above series can be written as a summation:

$$(6) \varphi = (e_{n+1})! - \big((a!\,i!)(e_{n+1})\big) - \sum_{a=2}^{e_{n+1}} \left(([a+2]!\,i!)\left(e_{n+1} - \frac{[a+2]}{2}\right)\right) \; where \; \varphi = legal \; arrangements$$

Given that a nondeterministic algorithm would be required to attempt all(until success was reached) possibilities, the upper bound for a given problem could be reduced through a deterministic approach.

## 8|An Outline of a Deterministic Solution

To solve contextualized problem, we begin by establishing sets of arrays, and defining the elements inside of them. *S*, is an array containing all elements of the system

$$I \in S \; ; \; length(S) \geq I$$

*Incompatibles,I,* is a list of all incompatible elements. Each pair is a binary system, meaning that at most, 1 element of S is incompatible with any given element of S. Therefore the S array length shall first be determined in the following manner:

$$\begin{matrix} name1 & 1 \\ name2 & 2 \\ name3 \to & 3 \\ name4 & 4 \\ \dots & \dots \end{matrix}$$

By following the above method all Incompatibles(each set of two) are given the same value. All $I \in S$ as previously mentioned therefore all *Incompatibles* must be eliminated from S. Also the last number used in *Incompatibles* shall be stored in a variable; a. A new $2 \cdot n$ array is the D list and in it, within column 1 only, the list of incompatibles shall be placed:

$$\begin{bmatrix} name1 & - \\ name2 & - \\ \dots & - \\ \dots & - \end{bmatrix}$$

Currently $a \cdot 2 = length(I) = length(D)$. However before the $D$ list can consist of more than simply $I$, all $I$ must be eliminated from $S$.

All elements of I must be copied to an array ι and all elements of the first row of I shall be copied from I to D and then deleted from I and S. Once the length of I is equal to 0, remaining elements of S are placed in the second column of D until the length of D is greater than the length of $\frac{I}{2}$. Once $\frac{I}{2}$ is greater than D, the remaining elements of S are placed in D in both the first and second columns.

## 9| Algebraic Program

## Notation

The following bi-conditional notation will be used frequently throughout this Algebraic Program: $X \& Y \Rightarrow < M|\xi^*Y >$ states that given X is true and Y is true allow the machine M to perform ξ on language Y.

O*:X means that the operation O is defined as X

Matrices without brackets such as: $\frac{I_X}{I_Y}$ which appear next to a bracketed matrix are not actual matrices, rather they are intended to denote row numbers.

\ is a special form of and, it is not the same as the Boolean and or ∧, rather it means that X\Y means that X is performed and Y is performed regardless of their condition

Semicolons denote the change from one statement to the next

$ is the length (Rows) of a given matrix: $X$ returns the length of X

## Tripound Algorithm

In order to create a Boolean Program for the solution presented previously we must define variables: $line \in Integers$, S is the list of all elements, *Incompatibles* abbreviated as $I$ is a N×2 matrix which contains incompatibles for a given element of S in the form:

$$I_X \begin{bmatrix} X & Y \\ Y & X \end{bmatrix}$$
$$I_Y$$

Which states that $\{X, Y\} \in I_X \leftrightarrow \{X, Y\} \in I_Y$. Note the input need not be in order of X and Y reversals, this will be clear from the program.

$$k = \$S\$$$

$$line = 0$$

I=ι ; $I_X \in i \; \wedge \; I_X = i_X$

(a) $\forall (line < k) \Rightarrow \; < M | dD >; d : D_{(line \;\; 1)} = I_{[line \;\; 1]}$

$$line = 0$$

(b) $\forall (line < k) \& (S_{line} \in I) \Rightarrow \; < M | v^* S > \backslash k = k + 1 \backslash < M | hI >; h : I - H; H = I_Y \begin{pmatrix} X & Y \\ Y & X \\ 0 & 0 \end{pmatrix} \; ; \; v^* : S - A; A = I_Y \begin{pmatrix} X \\ Y \\ 0 \end{pmatrix}$

$I_X$ on H and A labels, $I_n$ labels.

$$line = 0$$

(c) $\forall \left( line < \frac{k}{2} \right) \& \left( line \leq \frac{\$\iota\$}{2} \right) \Rightarrow \; < M \left| KS >; K : S_{line} = D_{(line \;\; 2)}; \forall \left( line < \frac{k}{2} \right) \oplus (line \leq \$\iota\$) \Rightarrow \; < M \right| JS >$

$$J: D_{(line\ 1)} = S_{line} \backslash line = line + 1 \backslash D_{(line\ 2)} = S_{line} \backslash line = line + line; halt\ if\ line > \frac{k}{2}$$

# 10|Time Complexity for the proposed Algorithm

**Definition**

A multiplicative compound algorithm is the set of sub-algorithms W' which constitute the major Algorithm W in such a way that each sub-algorithm has steps which are effected by previous time limits such as the classic schoolbook long multiplication algorithm which requires a multiplication step which is followed by addition for each integer.

An additive compound Algorithm is the set of sub –algorithms W' which constitute an Algorithm W in such a way that each sub-algorithm simply adds to the time of the previous upper limit. An example of such a compound algorithm would be the addition of multiple numbers separately.

**Time Complexity Analysis of the Tripound Algorithm**

The combination of (a),(b),and (c) create a compound algorithm, denoted as ABC. All algorithms within ABC are addition and Subtraction operations performed on matrices.

$k = \$S\$$ Determines the length of S, this operation is O(n)

$line = 0$ declaring a variable, essentially 0





I=$\iota$ ; $I_X \in i \;\wedge\; I_X = i_X$ Copying one matrix to another is simply O(n)

**Algorithm (a) Individual Maximum**

$$\text{(a)} \quad \forall (line < k) \Rightarrow\; <M|dD>; d\colon D_{(line\;\;1)} = I_{[line\;\;1]}$$

(a) Performs a matrix copying operation for each element and then a subsequent pasting

operation for half therefore it is O(($O$)$^2$).

**(a) Output Effects on Later Functions**

(a) does not affect (b) or (c) because (a) does not add or subtract to a language of (b) or (c).

**Algorithm (b) Individual Maximum**

(b) $\forall (line<k)\&(S_{line}\in I)\Rightarrow\;<M|v^*S>\backslash k=k+1\backslash <M|hI>; h\colon I-H; H=I_Y\begin{array}{c}I_X\\Y\\I_n\end{array}\begin{pmatrix}X&Y\\Y&X\\0&0\end{pmatrix}\;;\;\;v^*\colon S-A; A=I_Y\begin{array}{c}I_X\\Y\\I_n\end{array}\begin{pmatrix}X\\Y\\0\end{pmatrix}$

$<M|hI>$ is an operation which reduces the I matrix to reduce the search area of the ($S_{line} \in I$)

search operation. $<M|hI>$ simply reduces a function which is already linear O(n). Therefore

the time for the search becomes progressively smaller as each loop is performed.

$<M|v^*S>$ is a matrix subtractive operation which reduces the length of to :$S$-$I$=U.

k=k+1 is O(n) as it is an addition operation[2]

**(b) Output Effects on Later Functions**

$<M|v^*S>$ reduces $S$ by $I$ thus $S$-$I$=U therefore operations of S are controlled by the

length U within Algorithm (c).





**Algorithm (c) Individual Maximum**

$$(c) \quad \forall \left( line < \tfrac{k}{2} \right) \& \left( line \leq \tfrac{\$t\$}{2} \right) \Rightarrow < M \left| KS >; K: S_{line} = D_{(line \quad 2)}; \forall \left( line < \tfrac{k}{2} \right) \oplus (line \leq \$t\$) \Rightarrow < M \left| JS >$$

$$J: D_{(line \quad 1)} = S_{line} \backslash line = line + 1 \backslash D_{(line \quad 2)} = S_{line} \backslash line = line + line; halt\ if\ line > \frac{k}{2}$$

<M|KS> is an arbitrary copying and pasting operation : $O\left(\frac{U^2}{4}\right)$

<M|JS> is also a copying operation which is simply of time complexity: $O\left(\frac{U^2}{4}\right)$

**Summary:**

$\frac{n}{c} = U$ where c is an arbitrary constant. Therefore it follows that by multiplication and addition the time complexity of the system must be of time $O(c_1 n^k + c_2 n^{k-x} \dots)$ and therefore the compound Algorithm has a maximal run time which is measured by a polynomial.

# 11|Determinism

Deterministic Algorithms are those which operate in the same manner given an input I at any time T. The Tripound Algorithm is simply a set of matrix operations none of which overlap and none of which depend on external input. Therefore by inverse corollary it follows that the Tripound Algorithm is Deterministic algorithm.

# 12| Time Complexity Minimization

$I \in S$ therefore the searching algorithm is required to find all of $I$ within $S$. However, if the Incompatibles were listed at the top of $S$ then the algorithm could simply delete from $lines=0$ to $lines=a$ of $S$.

## REFERENCES

1.  Cook, Stephen A. "The complexity of theorem-proving procedures." In *Proceedings of the third annual ACM symposium on Theory of computing*, pp. 151-158. ACM, 1971.

2.  "P vs NP Problem." The Clay Mathematics Institute. Accessed June 21, 2014.

    http://www.claymath.org/millenium-problems/p-vs-np-problem.

3.  Ignjatovic, A., Alex North, and David Greenaway. "Complexity of Basic Operations on Natural Numbers." UNSW INDEX. Accessed June 24, 2014.

    http://cgi.cse.unsw.edu.au/~cs3121/Lectures/.

4.  "Non-deterministic Algorithms." Non-deterministic Algorithms. Accessed June 24, 2014.

    http://cs.nyu.edu/courses/spring03/G22.2560-001/nondet.html.

5.  Cook, Stephen. *The P Versus NP Problem*. Technical paper.